# Chatbots as conversational healthcare services

**Mladan Jovanović**
Singidunum University, Belgrade, Serbia

**Marcos Baez**
LIRIS - Claude Bernard University Lyon 1, France

**Fabio Casati**
University of Trento, Italy
ServiceNow, Santa Clara, US

*Abstract*—Chatbots are emerging as a promising platform for accessing and delivering healthcare services. The evidence is in the growing number of publicly available chatbots aiming at taking an active role in the provision of prevention, diagnosis, and treatment services. This article takes a closer look at how these emerging chatbots address design aspects relevant to healthcare service provision, emphasizing the Human-AI interaction aspects and the transparency in AI automation and decision making.

■ **CONVERSATIONAL** systems are entering our everyday lives, such as Amazon Alexa and Google Assistant. Beyond such all-in-one systems, there are growing demands for building conversational services in healthcare. The services are using a shared design metaphor - a personal assistant that provides healthcare through natural conversation. The main reason is making online healthcare more user-friendly - an agent takes a patient through a turn-taking dialog, similar to how doctors do [1], [2].

The recent transformation of digital healthcare aims at providing personalized health services and helping patients in self-managing their conditions [1]. Chatbots are becoming part of this paradigm shift as a cost-effective means to deliver such services [1]. Besides, they facilitate well-being as ingraining positive self-care habits [1]. The main benefits are ease of use and accessibility - the conversation metaphor makes them more intuitive, available on smartphones everywhere, anytime [3]. However, healthcare provision still depends on health professionals. Therefore, inspiring researchers and practitioners to explore the potential of Conversational AI to bring personalized services through automation [1].

The growing number of healthcare chatbots, partly due to the democratization of chatbot development, motivates a closer look at how the systems address aspects concerning user experience, adoption and trust in automation, and healthcare provision. While a recent literature review provides insights on general design considerations for healthcare chatbots [4], we focus on publicly available chatbots and take a more domain-specific approach in identifying relevant design dimensions considering the specific role in healthcare provision.

We report on a systematic analysis of 158 publicly available healthcare chatbots. Due to their very nature, we believe that the form and function of the healthcare chatbots cannot be neatly separated and are equally important. This paper:
- identifies salient service provision archetypes that characterize the emerging roles and functions the chatbots aim to fulfill;
- assesses the design choices concerning domain-specific dimensions associated with health service provision and user experience;
- provides implications for theory and practice that highlight existing gaps.

## ANALYTICAL FRAMEWORK

A healthcare chatbot can be conceptualized as a set of interconnected layers. The *knowledge layer* contains the domain and user databases. The information from this layer is an input for the *service layer* of healthcare provision. This layer implements healthcare decision-making processes. Once it generates the decisions, they are communicated to a *dialog layer*. The rule-matching dialogs are robust and straightforward to build but work in a constrained domain, whereas probabilistic (machine learning) tools may provide more natural dialog but lack robustness [5]. The dialog layer extracts user intentions, creates responses by consulting the service layer, and communicates them to the *presentation layer* that implements a text- or voice-based UI.

We introduce the analytical framework containing the attributes to characterize and compare existing healthcare chatbots. The framework captures the domain-specific aspects of healthcare provision, emphasizing the Human-AI interaction aspects and the transparency in AI automation and decision making. The dimensions are summarized below and detailed in **Table 1**:

- **Conversational style.** While deploying suitable and successful dialog strategies is still an open challenge in Human-AI interactions, some domain-specific dimensions emerge from research in health information systems [6] and recent general guidelines for Human-AI interaction [7]. Sociability, empathy, understandable medical vocabulary, and emerging dialog styles are among the key design dimensions we analyze.

- **Understanding users.** The users' ability to express intentions and be understood by the chatbots is another fundamental challenge in dialog-based interactions. Data collection methods (explicit or implicit) and the ability to recover from conversation breakdowns are among the critical functions of healthcare, shaping user's expectations of natural dialog capabilities.

- **Accountability.** There are ethical and practical reasons for making AI more transparent and explainable. Not only concerning model biases and privacy concerns but also to understand the reasoning behind algorithmic decisions that could have a significant impact on healthcare service provisioning [8]. The implementation of these features can address privacy concerns, build trust, and make the service provisioning more accountable for users [8].

- **Healthcare provision.** The domain-specific aspects of service provisioning include the type of the chatbots' role, emerging functional archetypes within these roles, collaboration facilitated by the chatbot, and continuity of service delivery.

Using the analytical framework, we identified and characterized publicly available healthcare chatbots in the English language, as of August 2020. Starting from the health provisioning roles, we analyzed how the other chatbot design dimensions are implemented for the primary functions. To this end, we screened health-related chatbots from two popular databases, Botlist (https://botlist.co) and Chatbots.org (https://chatbots.org) in the categories "Health and Fitness" and "Body health". We included other available, well-known examples that are often analyzed in the scientific literature and appear at the top search results when searching for chatbots for health. Our list is a representative sample with a clear overview of the chatbots' use. A total of 225 chatbots were screened and annotated by health provisioning roles by two researchers, resulting in 158 relevant health chatbots (coding agreement 90%). Two researchers independently annotated the chatbots' functions in an emergent coding scheme, which was then consolidated by consensus to describe the salient archetypes for each role. The analysis focused on archetypes that describe a direct involvement of the chatbot in the healthcare service provisioning, emulating the functions of a healthcare professional (e.g., in performing a diagnosis, or delivering a therapy). It was the case for the 6 of the 9 archetypes identified. We excluded chatbots from the archetypes "support for diagnosis", "access to healthcare" and "support for therapy", where chatbots act as mediators to facilitate access to healthcare services, information and products. Accordingly, we had around 7-8 top chatbots from 6 archetypes, for a total of 45 chatbots. We selected popularity (e.g., number of views and likes) as a measure of quality and adoption, thus focu-sing the analysis on the most widely adopted chatbots. A scoring system was derived to complement the qualitative observations, and describe the level of implementation of each dimension: Low, indicating that the dimension was not explicitly addressed (e.g., explainability: the chatbot does not provide any explanations for its decisions); Medium, showing partial implementation (e.g., explainability: the chatbot provides some evidence, but important details are still missing); High, meaning a high degree of implementation (e.g., explainability: the chatbot explains major decisions). The supplementary material including details about the process, scoring system, dialog examples and resulting annotated dataset is available at https://cutt.ly/TdOzKPm.

Table 1. Analytical framework.

| Dimension | Attribute |
|---|---|
| Conversational style | **Sociability.** Social communication (i.e., small talk) is critical for sustained user engagement [9], a premise of the effectiveness of a health intervention. As a property (content) of the conversation itself, it builds and maintains social bonds among interactants [10]. In this regard, we look at whether chatbots implement social conversation capabilities. |
| | **Empathy.** Another desirable characteristic of chatbots is exposing empathy, the ability to recognize users' emotions and respond appropriately to the current mood [1], [3], [9], [11], and even more so in vulnerable scenarios posed by health services. Thus, we qualitatively assess if chatbot dialogs provide empathy cues in their conversations. |
| | **Vocabulary.** Adapting the conversation content to a suitable and understandable medical vocabulary is also important for the quality of the healthcare provision [1]. We analyze strategies and features adopted by chatbots to address this aspect explicitly. |
| | **Proactivity.** A mix of *proactive* and *reactive behavior* is another inherent feature of everyday human communication that AI aims to replicate [7], and that can inform how services are provided. We examine whether chatbots display proactive behaviors in providing their services. |
| Understanding users | **Data collection.** An important aspect is understanding the input patterns and *data collection* methods enabled by chatbots as they inevitably balance the robustness and naturalness of conversations [5]. In this regard, we qualitatively assess emerging input patterns, and determine whether the chatbots leverage on explicit and implicit data collection strategies. |
| | **Error recovery.** *Error recovery* strategies are crucial for addressing the breakdowns and preventing from degrading the user experience and drawing incorrect decisions [12]. We assess whether chatbots implement error recovery strategies, focusing on the ability to deal with human error. |
| Accountability | **Explainability.** We define explainability as the ability of the chatbot to inform and explain its decisions (e.g., how a diagnosis was reached, or why an activity program was changed). |
| | **Transparency.** We look at the transparency with regard to data collection practices (e.g., why is the chatbot collecting certain information). |
| Healthcare provision | **Role.** It indicates the chatbot's role(s) in healthcare provision as diagnosis, prevention, and therapy. Some chatbots may play multiple roles. |
| | **Archetype**. It describes emerging service patterns within the health provision role. |
| | **Collaboration.** Together with proper integration with healthcare infrastructure as a means for augmenting skills of medical professionals [3]. When analyzing collaboration, we focus on identifying the stakeholders involved and the type of technology-mediated interactions enabled by the chatbot. |
| | **Continuity.** Refers to the time of the service delivery, whether in *one-time* sessions (akin short-term visits) or leveraging on the opportunity for more *continuous* healthcare delivery [1], [2]. |

We detail our analysis in the following sections, describing the emerging archetypes and salient design features as characterized by our framework.

CHATBOTS FOR DIAGNOSIS

Diagnostic chatbots check user's symptoms and recommend courses of action. Three general archetypes of diagnosis chatbots emerged from our analysis:

- **Support for diagnosis** (10/32). The archetype does not perform the diagnosis but instead support a diagnosis by either i) facilitating access to health services, such as the Pathology Lab Chatbot facilitating access to doctors and scheduling visits, ii) supporting online consultations with health professionals, such as the iCliniq that pairs up users with doctors for online consultation, and iii) providing conversational access to information regarding symptoms and diseases, such as the WebMD.

- **General symptom checker** (15/32). The archetype is mimicking a consultation with a general health professional, walking users through a series of questions regarding their symptoms to diagnose a condition, and, in some cases, suggests a course of action. A prominent example is HealthTap, a chatbot that collects symptoms and provides potential causes in dialog-based interactions.

- **Specific symptom checker** (7/32). This archetype aims at either i) helping users confirm the presence and severity of an ailment, or ii) diagnosing a particular condition, akin to having a consultation with a medical specialist. An example from the first category is FeverBot, which helps users determine whether they require medical attention, and for the second, the Mental Care Bot, which specializes in diagnosing mental disorders.

The archetypes have different foci but follow a typical dialog structure, consisting of profiling the user, collecting and refining symptoms, diagnosis, and follow-up. This process is typically enacted in one-time sessions involving a user and the chatbot, not reusing previously collected information - even though some chatbots offer symptoms journaling (e.g., HealthTap).

Collecting and refining symptoms is approached with different dialog styles. Specifying symptoms in natural language (e.g., "I have back pain") has varying levels of success. The chatbots try to identify the symptom either directly from the user input (e.g, Your.MD), directing the user input to a search page (e.g., Babylonheath), or a combination of both. Follow-up questions to refine the symptoms (e.g., "Which part of your back is hurting") display a closed list of predefined options (e.g., "Lower back" or "Upper buttock area") requiring users to select an option from a list (e.g., Ada), or swiping through illustrated cards (e.g., HealthTap). The symptom checkers for skin problems (e.g., Skinive) have the possibility of uploading pictures to bootstrap the diagnosis, using computer vision to interpret the input. Interestingly, none of the chatbots make use of implicit data collection (e.g., sensor data), but collect user information explicitly during the conversations. Allowing users to edit and backtrack information is an error recovery mechanism absent in almost half of the chatbots (e.g., Buoyhealth provides an "Edit" option on each user input).

The majority of chatbots interact with users following a scripted interview without being cautious of the users' responses. Making technical language understandable is a strategy implemented explicitly by only three chatbots. They address this aspect by either including contextual help for each question (e.g., Ada: "What does it mean"), presenting pictures of the symptoms and options (e.g., HealthTap), and indirectly by requesting feedback on questions (e.g., Buoyhealth: "This is confusing"). Transparency clarifies potential privacy concerns regarding sensitive questions. Despite its importance, only one chatbot explicitly addressed this issue, namely Buoyhealth, allowing users to inquire about the reasons behind each question ("Why am I being asked this?").

The typical diagnosis report is a list of potential causes, explaining the reasoning behind and courses of action. It consists of the information describing: i) strength of the evidence supporting the diagnosis, ii) symptoms present for the cause, iii) type of care recommended and the specialist needed, and iv) possible actions. Except for three chatbots, the majority explains their decisions by providing evidence connecting reported symptoms to the potential causes. All chatbots explicitly inform users that the report does not replace a medical consultation.

CHATBOTS FOR PREVENTION

The chatbots in this role assist in tracking and building awareness of a user's health and help prevent health declines by building desirable habits. The prevention is offered through a range of services [1] that can be aggregated into three archetypes:

- **Access to healthcare** (12/109). The chatbots from this archetype do not participate in the provision of healthcare service but represent an entry point to using these services. Its main goal is to increase the efficiency of healthcare services by reducing the effort and increasing the speed of access. They do this by i) connecting patients to healthcare professionals, ii) discovering medical drugs online, or iii) providing healthcare customer service tasks. For example, the iClinic provides 24/7 medical customer service for patients, such as booking appointments with their doctors. The Project Alta facilitates the discovery and purchase of pills for improving cognitive functions.

- **Health education** (20/109). The educational archetypes prevent by teaching users on prevention procedures for specific health conditions. For instance, DoctorBot provides healthcare information on different topics. A very recent example is Jennifer, a chatbot designed to combat misinformation and answer questions on the COVID-19 virus.

- **Health coaching** (77/109). Its goal is to prevent health degradation by improving general well-being and inducing a healthy lifestyle. At its core are psychological incentives to maintain or facilitate desirable behaviors. Their functions can be categorized as i) personalized reminders for mental exercise and workouts, ii) psychological motivators for mental and physical practices, or iii) advisors on positive habits regarding sleep, nutrition, and well-being.

For example, the FitCircle uses reputation-based incentives for exercising (such as goal-setting and progress information), while the StopBreathe&Think recommends mental exercise for psychological well-being. The Forksy is a nutrition assistant who advises on nutriments tailored to the user's health goals and eating habits.

Concerning conversational style, the first archetype specializes in a specific task, such as connecting with a doctor, booking an appointment, or ordering particular medicine. The latter two are flexible in a sense that they educate on related, but different topics, or coach on a range of well-being activities within a type or across several types (physical, mental, nutritional). All archetypes offer instrumental, goal-oriented conversations in which they guide users through predefined programs of exercises. Regarding the vocabulary, less than half the chatbots referred the users to external glossaries of terms for additional explanations. The archetypes employ proactive conversation during prevention, by probing users for necessary information. The user information is collected explicitly, from the conversations. The error recovery is present as either asking the users to rephrase the misunderstood input or jumping to the beginning of the dialogs.

Regarding accountability, the archetypes do little to explain the specific decision to the users (e.g., general explanations on their websites). The transparency with the archetypes is low in the sense of not highlighting nor clarifying the reasons for collecting data from their users.

The chatbots from the first archetype follow a scripted, question-answering dialog flow without keeping conversation history. For example, Healthy Recipe HQ implements on-demand, dynamic question answering by aggregating available information online. The second archetype implements a similar dialog structure while educating on a specific topic. The third archetype offers a more complex dialog with flexible conversation based on the history and user profiling. Health coaching chatbots recommend actions based on user monitoring through sustained conversation. The actions originate from activities' pools including workouts, nutrition plans, and mental exercise programs. They continuously motivate users for a healthy lifestyle by combining psychological incentives that include self-reflection on achieved progress, reminders for activities, and the evidence from peers. Some health coaching chatbots connect with peers, such as FitWell. Otherwise, the majority targets individuals.

CHATBOTS FOR THERAPY

The role assists or provides treatment of specific health declines or conditions (such as pregnancy or therapeutic diet). The therapy services can be grouped into the following archetypes:

- **Support for therapy** (7/41). This archetype assists during the phases of the treatment. The examples are personalized reminders to medication adherence as part of the therapy (e.g., Florence), or listing medicines based on positive online user reviews for natural health cures (e.g., HealthRobot).

- **Health therapy** (20/41). The therapist archetype takes a more active role by providing at-home therapy for its patients. Based on their primary target, they offer either i) drug-based therapy or ii) practice-based therapy for its patients. The first sub-archetype recommends and tracks medicine use during the treatment (such as Florence). The second sub-archetype provides practical guidance on the activities for successful treatment. For instance, KetoBot suggests a ketogenic diet to fight against diabetes.

- **Cognitive behavioral therapy (CBT)** (14/41). This archetype provides a range of therapies that target specific mental states and emotions. The therapy is a structured, guided conversation that starts with a question-answering to identify the patients' condition. It continues by recommending specific exercises based on the estimated conditions and tracking the target state. The measures of treatment's progress are self-reported, provided by patients as a free-form text. Woebot is a personalized mental therapist who tracks users' mood and suggests mental activities. Wysa aims at improving patients' mental health by providing emotional support. The common goal is to build resilience to mental disorders (i.e., stress, depression and anxiety) by developing positive habits (i.e., self-awareness and optimism).

The archetypes support multiple activities (i.e., facilitating access to different types of medicines) or health conditions (i.e., aiming at various health conditions). The CBT archetypes try to understand and respond to the users' current mood. This aspect is entangled with social elements, such as engaging in small talk on non-treatment topics. It increases the amount of conversation, specifically user-provided data, to improve the accuracy of guessing the user's emotions. As for the therapy-specific terms, the chatbots offer explanations during the conversations. The first archetype induces conversations through personalized reminders, whereas CBT chatbots initiate the dialogs on a time basis. The error recovery strategies include restarting current conversation, or asking additional questions for mutual understanding. User data are collected explicitly, from user input during conversations.

Concerning accountability, the minority of the therapy chatbots explain their decision to users, and clarify the reasons for collecting specific user data.

The supportive archetype uses rule-based or statistical approaches to dialog management. The former follows a predesigned turn-taking conversation (i.e., Meditation master to alleviate stress and sleeplessness). The latter implements a more natural back-and-forth message exchange that uses context information to generate responses. The second archetype employs flexible, probabilistic dialogs that preserve the conversation context. It follows a conversation pattern in which patients are screened for their condition(s), and guided and monitored throughout the specific treatment. The dialog adapts to the treatment's progress. The third archetype follows a similar principle. The context information is extracted from user profiling and conversation history (i.e., Wysa in monitoring therapy progress). It focuses on multiple mental/emotional states and conversation as a means of therapy, offering more fluid dialogs (i.e., Woebot). Archetypes mainly target individual users.

IMPLICATIONS FOR THEORY AND PRACTICE

Current healthcare chatbots remain a supplementary service rather than a replacement of the medical professionals.

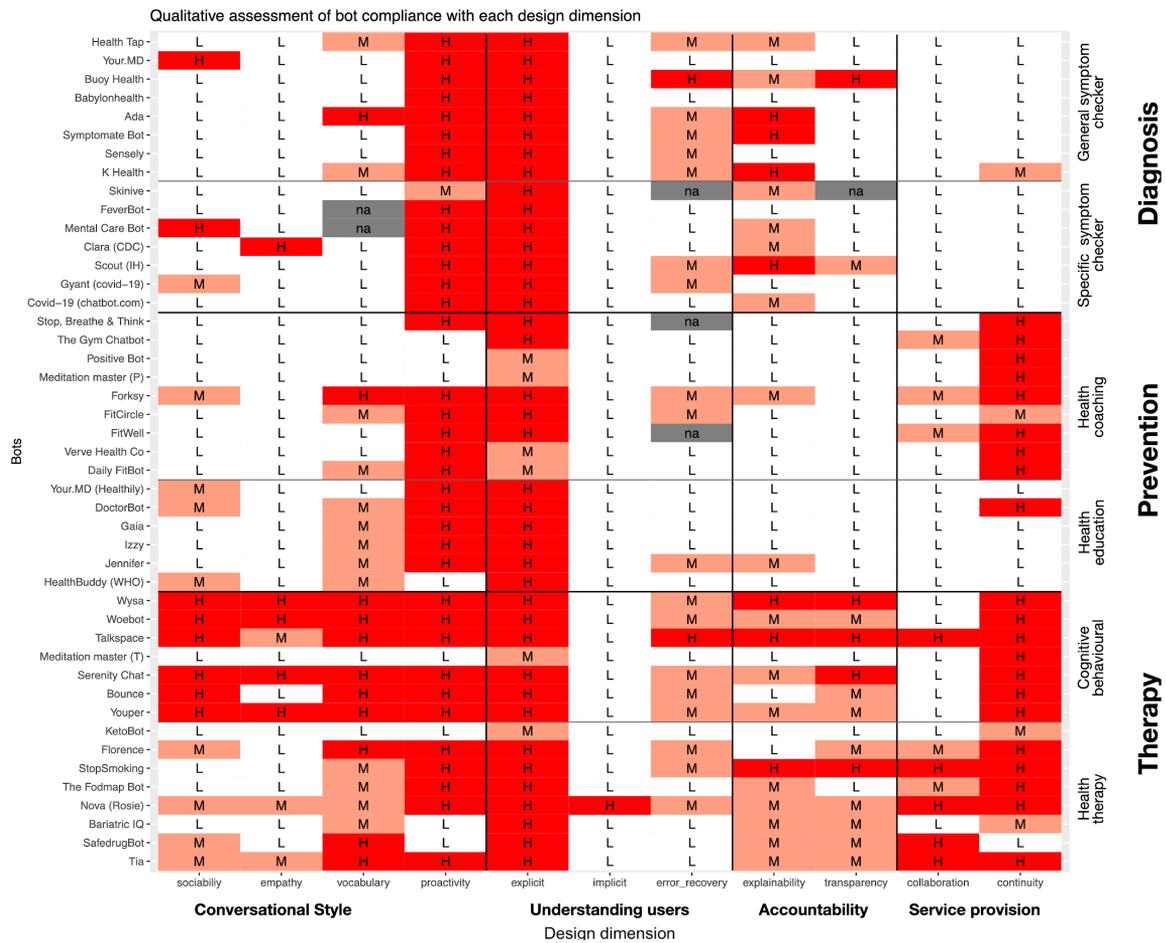

Figure 1. State of the healthcare chatbots regarding the analytical framework (L - low, M - medium, H - high).

The archetypes share levels of engagement as i) *an active role* in healthcare provision, emulating the functions of a healthcare professional; ii) *facilitating access* to healthcare services by matching users with service providers, or supporting the service delivery; and iii) *providing users* with information and products. Some chatbots expose multiple roles. For example, Florence instructs medicine intake during therapy, and provides information about a disease, for prevention.

*Conversational style.* Our findings confirm the existing evidence that a practical, task-oriented goal conversation is dominant with healthcare chatbots [1], [2]. Except for the CBT archetype, chatbots do little to understand human social and emotional cues. Concerning the use of medical vocabulary, our analysis revealed that the diagnostic and preventative chatbots use expert vocabulary in healthcare provisions and should improve on making the terminology understandable to their users. As for the dialog initiative, preventative and therapeutic archetypes are more proactive than diagnostic archetypes that react to the users' questions.

Future healthcare chatbots should improve the conversation's social and emotional aspects while adapting to users' health literacy. The lack of these aspects may create dissonance in user experience due to false expectations and lead to rejection [9], [11], [13].

*Understanding users.* The chatbots' natural language capabilities remain limited, leading to alternatives such as breaking down the dialog in multiple questions and constraining user choices to deal with this limitation. This is particularly the case in diagnosis chatbots or data collection tasks where the user cannot narrate their condition but instead is guided through questions. Preventative and therapeutic chatbots offer more fluent dialogs by learning about their users and reusing conversation context. In analyzing the chatbots, we also noticed repair strategies [14] not implemented correctly in the dialog, sometimes even in its most basic form, such as preventing users from modifying previous inputs. Another aspect we noticed is that the archetypes rely mostly on explicit user input from conversations. The exception are chatbots that accept pictures of affected skin areas for automatic processing.

The above reveals that restricting interaction through close-ended questions may expand the chatbot's understand of the users, but errors related to human input or incorrect references can still emerge in interactions. Improving error management is the emerging requirement for health chatbots. As for opportunities, chatbots can significantly benefit from continuous implicit data collection using smart devices' sensing technologies currently widespread [15].

*Accountability.* We discovered that *transparency* in data collection is insufficient in the archetypes. The minority of the analyzed chatbots give reasons for asking users for their data (Figure 1). Concerning *explainability*, the chatbots do little to provide causes or explain the reasons behind their healthcare decisions. These can raise concerns with users about chatbots' accountability.

The implications are two-fold. Firstly, public expectations of the chatbots need to be set explicitly, in advance. Secondly, there is a need for greater transparency and explainability of the logic behind each archetype. In particular, i) explaining why and how the chatbot collects certain information, and ii) clarifying all the relevant decisions taken during the service provision (e.g., why diagnosing an illness, prescribing a drug, or increasing the workout intensity).

*Healthcare provision.* Traditionally, the patient's condition is assessed during sporadic, short-term visits to healthcare facilities, and critical decisions may be affected by the uncertainty of the measurements. We observed that not many chatbots capitalize on the opportunities for *continuity* in the service provisioning. Diagnostic chatbots provide one-time sessions with little to no information shared across sessions. Therapeutic and health coaching archetypes offer prolonged, shared interactions with patients at their homes. Currently, healthcare chatbots are stand-alone applications, independent of healthcare systems. Concerning *collaboration*, the archetypes focus on individuals as a user-chatbot relation. Group dynamics are missing, supported by a handful of chatbots as user-chatbot-doctor relations (i.e., Sensely), or forming peer groups (i.e., FitWell).

The chatbots need to address the above aspects through continuous service and better integration, opportunities not entirely leveraged by current healthcare systems [1], [2]. The future healthcare chatbots should also engage and moderate among multiple actors consisting of patients, their social circles (family and friends), and caregivers. This assumes reusing social context by leveraging the humans and intelligent agents, the environment, and the healthcare infrastructure [16].

CONCLUSION

Healthcare chatbots are yet to capitalize on the opportunities provided by conversational media to provide better dialog-based interactions appropriate to the task, and with the social intelligence to manage interaction in potentially vulnerable scenarios. Our work provides the first step towards these goals by characterizing the emerging roles of chatbots in service provisioning and highlighting design aspects that require the community's attention. We believe our findings can guide researchers in identifying and validating dialog patterns appropriate to the existing archetypes, and practitioners in understanding the emerging use cases of chatbots in healthcare provision. Future work should focus on understanding the actual medical value of the chatbots and their effects on health outcomes and user experience.

**Limitations.** Our study does not attempt to evaluate an exhaustive list of existing health chatbots, but a representative sample of the current landscape. We acknowledge that the popularity metric used for selecting chatbots for full evaluation is an approximation, but given the number of chatbots evaluated any potential misrepresentation should be alleviated. The evaluation did not address the effects of chatbots on health outcomes or the evidence supporting the health service. Similarly, we did not analyze the content, nor assess it for accurate and evidence-based information. They are important aspects to be addressed in future work.

Mlađan Jovanović received PhD in Computer Science at the University of Belgrade in 2013. He is an Assistant Professor at Singidunum University, Belgrade. Before academia, he worked in the industry, focusing on the interactive computing systems. His main research interests include conversational user interfaces and human-centered AI. Contact him at mjovanovic@singidunum.ac.rs.

Marcos Baez is currently a Research Fellow with Claude Bernard University Lyon 1, Villeurbanne, France. His research interests include Web engineering, crowdsourcing and Human-AI interaction. He received the PhD degree in Computer Science from University of Trento, Trento, Italy. Contact him at marcos.baez@liris.cnrs.fr.

Fabio Casati is a principal machine learning researcher at ServiceNow. He is a Professor of Social Informatics at the University of Trento, where he started research on hybrid human-AI computations and technologies for happiness and life participation. He is the coauthor of a best-selling book on Web Services and author of over 200 peer-reviewed articles. Contact him at fabio.casati@unitn.it.